\documentclass[12pt]{iopart}
\usepackage{amsmath}
\usepackage{iopams}

\begin{document}

\maketitle

\title{Fully Symmetric Relativistic Quantum Mechanics and 
Its Physical Implications}

\author{B. D. Tran and Z. E. Musielak}
\address{Department of Physics, The University of Texas at 
Arlington, Arlington, TX 76019, USA}
\ead{bao.tran3@mavs.uta.edu; zmusielak@uta.edu}

\begin{abstract}
A new formulation of relativistic quantum mechanics is presented 
and applied to a free, massive, and spin zero elementary particle in the 
Minkowski spacetime.  The reformulation requires that time and space, 
as well as the timelike and spacelike intervals, are treated equally, which 
makes the new theory fully symmetric and consistent with the Special 
Theory of Relativity.  The theory correctly reproduces the classical action 
of a relativistic particle in the path integral formalism, and allows for the 
introduction of a new quantity called vector-mass, whose physical implications 
for nonlocality, the uncertainty principle, and quantum vacuum are described 
and discussed.
\end{abstract}


\section{Introduction} 

Relativistic Quantum Mechanics (RQM) primarily concerns free relativistic fields [1,2] described by 
the Klein-Gordon [3,4], Dirac [5], Proca [6] and Rarita-Schwinger [7] wave equations, whereas 
the fundamental interactions and their unification are considered by the gauge invariant Quantum 
Field Theory (QFT) [8,9].  The above equations of RQM are invariant with respect to all transformations 
that form the Poincar\'e group $\mathcal {P}\ =\ SO (3,1) \otimes_s T(3+1)$, where $SO(3,1)$ 
is a non-invariant Lorentz group of rotations and boosts and $T(3+1)$ an invariant subgroup 
of spacetime translations, and this structure includes reversal of parity and time [10].  Bargmann 
and Wigner [11] used the representations of $\mathcal {P}$ [12] to obtain the Bargmann-Wigner 
equations, which reduce to the RQM equations under special conditions.  Different methods of 
deriving the RQM equations were proposed by using either symmetry properties [13] or the
irreducible representations (irreps) of $T(3+1)$ and the resulting eigenvalue equations [14,15].

There were several attempts to generalize the RQM equations, specifically, the Klein-Gordon 
[16-19], Dirac [20-25], or both Klein-Gordon and Dirac [26-28] and other [29] equations. 
Physical motivations for these generalizations were different and they range from including 
some supersymmetry effects to unification of leptons and quarks and accounting for different 
masses of the three generations of elementary particles.  Recently, the Dirac equation was 
generalized to include chiral symmetry and it was shown that the correct identification 
of chiral bases may explain the small masses of neutrinos and be used to predict a new 
massive particle of dark matter [30].  

As required by Special Theory of Relativity (STR), time and space must be treated
equally, which means that all basic equations of RQM are symmetric in their time
and space derivatives.  In STR, the timelike and spacelike intervals are also equally
allowed, even so RQM has been formulated using the timelike intervals only [1,2]; 
known exceptions are some tachyonic versions of RQM and QFT [31,32].  The main 
objective of this paper is to reformulate RQM by taking into account both timelike 
and spacelike intervals and explore physical implications of the developed theory. 
The presented theory is developed for a free, massive, and spin zero elementary 
particle in the Minkowski spacetime.  We follow Wigner [12] and define an 
elementary (quantum) particle as an object whose wavefunction transforms as 
one of the irreps of $\mathcal {P}$ [10]. 

The presence of the spacelike intervals in our theory modifies its linear momentum
that, in addition to its timelike eigenvalues, has also spacelike eigenvalues.  As a 
result, the STR energy-momentum relationship with its scalar-mass is modified by 
an additional term with its vector-mass, which only affects the spatial distribution
of particle's wavefunction.  The physical consequences of the presence of the 
vector-mass in the energy-momentum relationship are far reaching, starting with
the modifications to the mass term in the Klein-Gordon equation, and then leading
to nonlocality, some modifications of the uncertainty principle, and a novel picture
of vacuum.  Moreover, the developed RQM theory preserves causality, is self-consistent, 
and is formulated using the path integral formalism, which correctly reproduces the 
classical action of a relativistic particle; however, the theory does not allow for 
tachyons [31,32].  The theory is not a hidden-variable theory (e.g., [33]) and its 
physical interpretation is consistent with the Copenhagen interpretation of Quantum 
Mechanics (QM) [34,35].  

The paper is structured as follows: the generalized eigenvalue equations are 
presented and discussed in Section 2; the basic postulates used to reformulate
RQM are described in Section 3; the generalized Klein-Gordon equation and 
its solutions are given in Section 4; in Section 5, completeness of Hilbert 
space and causality of the theory are demonstrated; in Section 6, the theory
is formulated by using the path integral approach; physical implications of 
the theory are given in Section 7; and conclusions are in Section 8.

\section{Generalized eigenvalue equations}

The basis of our reformulation of RQM is a set of eigenvalue equations for the 
four-momentum operator $\mathcal{K}_{\mu} = i \partial_{\mu}$ acting 
on a relativistic quantum state $|\phi\rangle$ in the Minkowski spacetime, 
with the metric tensor $\eta_{\mu \nu}$ of positive time signature. The 
eigenvalue equations are given by 
    \begin{equation}
        \mathcal{K}_{\mu}|\phi\rangle=\left(P_{\mu} + V_{\mu}\right)\vert\phi\rangle\ ,
        \label{S2eq1}
    \end{equation}
where $P_{\mu}$ and $V_{\mu}$ are timelike and spacelike eigenvalues of 
$\mathcal{K}_{\mu}$, respectively, and $P_\mu P^\mu=m^2$, $V_\mu 
V^\mu=-\tilde{m}^2$ and $P_\mu V^\mu=0$, with $m$ being a scalar
mass and $\tilde{m}$ being a new quantity called here  "vector-mass" (see 
Section 3 for discussion).  The eigenfunction $\phi$ can be either scalar, 
or vector, or any higher rank tensor, or spinor wavefunction in order to 
be consistent with the irreps of $\mathcal {P}$ [12,10].  

As shown previously [14,15], the eigenvalue equations for $\phi$ being 
a scalar wavefunction represent the necessary conditions that this function 
transforms as one of the irreps of $T(3+1) \in \mathcal {P}$; here equation 
(\ref{S2eq1}) is a generalization of the previous results, which are 
recovered when $V_{\mu} = 0$.  A new RQM is formulated by using 
the above eigenvalue equations, which means that time and space, as 
well as the timelike ($P_{\mu}$) and spacelike ($V_{\mu}$) intervals, 
are treated equally as required by STR.

In this paper, we consider only {\it scalar wavefunctions} and follow
Wigner [10,12] to identify an elementary particle with one of the irreps
of $\mathcal {P}$.  In other words, the scalar wavefunction $\phi$
represents an elementary particle if, and only if, the wavefunction 
satisfies the above generalized eigenvalue equations.

The eigenvalue equations given by Eq. (\ref{S2eq1}) are expressed in 
terms of a general coordinate basis that is not necessarily orthonormal. 
Therefore, we formulate our theory in the local orthonormal basis, called 
tetrads, which are shown to be of great advantage in the development 
of our new RQM.  The tetrads are defined as local coordinate transformations 
from a global curvilinear coordinate system to a locally flat orthogonal 
coordinate system that is considered as the local inertial frame of an
observer in the Minkowski spacetime. 

We used Greek indices for global coordinate components and Latin indices 
to denote the components of local tetrad coordinates with $\eta_{ab}=
g_{\mu\nu}U_a^\mu U_b^\nu$ and $g_{\mu \nu}$ being 
the metric tensor of the global coordinate system. The tetrads are 
defined as $U^a_\mu=\frac{\partial\xi^a}{\partial x^\mu}$, where 
$\xi^a$ is the local coordinate function.  Then, the invariant spacetime 
element $\mathrm{d}s^2$ of the Minkowski metric and the eignevalue 
equations in the new local tetrad basis can be written as
    \begin{equation}
        \mathrm{d}s^2=g_{\mu\nu}\mathrm{d}x^\mu\mathrm{d}x^\nu=
        \eta_{ab}\mathrm{d}\xi^a\mathrm{d}\xi^b\ ,
        \label{S2eq2}
    \end{equation}
and
    \begin{equation}
        \mathcal{K}_a\vert\phi\rangle=\left(P_a+V_a\right)\vert\phi\rangle=
        K_a\vert\phi\rangle\ .
        \label{S2eq3}
    \end{equation}
We express the contravariant components of the four-vectors $P$ and $V$ 
in the local basis as $P=\{E,\boldsymbol{P}\}$ and $V=\{U,\boldsymbol{V}\}$ 
that locally satisfy the following conditions: $P_a P^a = 
E^2-\boldsymbol{P}\cdot\boldsymbol{P}=m^2$, $V_a V^a = U^2-
\boldsymbol{V}\cdot\boldsymbol{V}=-\tilde{m}^2$ and $P_a V^a =0$,
where $m$ and $\tilde{m}$ are the so-called scalar and vector mass, 
respectively.  It must be noted that both $m$ and $\tilde{m}$ are 
Lorentz-invariant constants, and that $U=\boldsymbol{P}\cdot\boldsymbol{V}/E$.

Let $\boldsymbol{\beta}=\boldsymbol{P}/E$ be the relativistic velocity of 
an elementary particle, which allows writing $P =\{E,\boldsymbol{\beta}E\}$, 
$V = \{\boldsymbol{\beta}\cdot\boldsymbol{V},\boldsymbol{V}\}$, and
$K=P+V=\{E+\boldsymbol{\beta}\cdot\boldsymbol{V},\boldsymbol{V}+
\boldsymbol{\beta}E\}$.  We refer to the vector $K$ as the four-mass vector
and use it to define the eigenstate wavefunction of a relativistic quantum 
particle.  The four-vector $P$ provides the energy-momentum relationship 
of STR, which is used in standard RQM.  However, the presence of the 
four-vector $V$ modifies this relationship to become
    \begin{equation}
        \left(E^2-\vert\boldsymbol{P}\vert^2\right)+\left(U^2-\vert\boldsymbol{V}
        \vert^2\right) = m^2-\tilde{m}^2=M^2\ ,
        \label{S2eq4}
    \end{equation}
which is the energy-momentum relationship for a massive and spin zero quantum 
particle used in the reformulated RQM.  This energy-momentum relationship 
reduces to that of STR, which is the basis for standard RQM, when $V = 0$.

It must be also noted that $M^2$ is not characterized as the mass of the particle, 
as in this reformulated RQM, a particle of definite mass and position in space must 
be in a superposition of different vector-mass $\tilde{m}$, and that the 
scalar-mass $m$ represents is responsible for the particle's world-line to be 
confined within the light cone.  On the other hand, the vector-mass 
$\tilde{m}$ does not play any role in the motion of the particle, even 
if its squared magnitude exceeds that of the scalar-mass, because of its spacelike 
origin.  Therefore, the theory developed in this paper only allows for particles with 
positive $m^2$ term, which rules out tachyons and tachyonic fields [31,32] as 
unphysical within the framework of this theory.

The origin of the four-vector $V$ is the generalized eigenvalue equations given by
Eq. (\ref{S2eq1}), and the above results show that the vector-mass $\tilde m$ is
associated with the presence of $V$.  In our reformulated RQM, $m$ is responsible
for the time evolution of the wavefunction $\phi$, like in standard RQM, while 
$\tilde m$ affects only the spatial distribution of the wavefunction $\phi$, which 
makes our reformulated RQM to be a fully symmetric theory that accounts for both 
the timelike and spacelike intervals.  Let us now describe the basic postulates of 
our new theory.  

\section{Basic postulates of new theory}

Our reformulations of RQM is based on the following postulates:
\begin{enumerate}
\item The canonical commutation relation of the conjugate operators 
$(\xi^a,\mathcal{K}_b)$ is given as
	\begin{equation}
        \left[\xi^a,\mathcal{K}_b\right]=-i\delta^{a}_{b}\ .
        \label{S3eq1}
	\end{equation}
\item The particle's relativistic three-velocity, $\boldsymbol{\beta}=\boldsymbol{P}/E$, 
is the boost parameter of a proper Lorentz transformation
	\begin{equation}
        \Lambda_{\boldsymbol{\beta}}=
        \begin{pmatrix}
        \gamma & \gamma\boldsymbol{\beta}^{\top} \\
        \gamma\boldsymbol{\beta} & \gamma\mathrm{I}+\frac{\gamma^2}{1+\gamma}\mathrm{B}^2
        \end{pmatrix}\ ,
        \label{S3eq2}
	\end{equation}
where $\boldsymbol{\beta}^\top$ is a row vector, the symbol $\top$ denotes 
matrix transposition, the $3\times 3$ identity matrix is represented by $\mathrm{I}$, 
$\gamma = \left(1-\vert\boldsymbol{\beta}\vert^2\right)^{-1/2}$ is the Lorentz factor
with $\vert\boldsymbol{\beta}\vert<1$, and
    \begin{equation}
        \mathrm{B}^2=\boldsymbol{\beta}\boldsymbol{\beta}^{\top}
        -\vert\boldsymbol{\beta}\vert^2\mathrm{I}
    \end{equation}
with
	\begin{equation}
        \mathrm{B}=\begin{pmatrix}
        0 & \beta_z & -\beta_y \\
        -\beta_z & 0 & \beta_x \\
        \beta_y & -\beta_x & 0
        \end{pmatrix}\ .
        \label{S3eq3}
	\end{equation}
From this, we have $E=\gamma m$ as the usual definition for the energy of a particle of 
scalar-mass $m$, and the new energy contribution related to vector-mass $\boldsymbol{k}$ 
is defined as $\boldsymbol{V}=\mathrm{L}\boldsymbol{k}$, where $\mathrm{L}$ is 
given by
    \begin{equation}
        \mathrm{L}=\gamma\mathrm{I}+\frac{\gamma^2}{1+\gamma}\mathrm{B}^2=
\mathrm{I}+\frac{\gamma^2}{1+\gamma}\boldsymbol{\beta}\boldsymbol{\beta}^\top\ .
    \end{equation}
Thus, the four-mass vector in the rest frame, i.e. $\boldsymbol{\beta}=\boldsymbol{0}$, 
is simply $K=k=\{m,\boldsymbol{k}\}$.

\item Let $\mathcal{U}_{\boldsymbol{\beta}}$ be an operator identified as a unitary 
representation of the Lorentz boost that acts on the state vector $\vert k\rangle=\vert m,
\boldsymbol{k\rangle}$ in a Hilbert space
	\begin{equation}
        \vert m, \boldsymbol{k};\boldsymbol{\beta}\rangle = \mathcal{U}_{\boldsymbol{\beta}}\vert 
        m,\boldsymbol{k}\rangle\ .
        \label{S3eq4}
	\end{equation}
Here, the states $\vert m,\boldsymbol{k}\rangle$ are the so-called inertial states describing 
a particle at rest, with $m$ and $\boldsymbol{k}$ forming the four-mass vector $K$, and 
the velocity $\boldsymbol{\beta}=\boldsymbol{0}$.  A particle can be boosted to a moving 
frame with $\boldsymbol{\beta}\neq\boldsymbol{0}$, so the state of a moving particle is 
labeled as $\vert m,\boldsymbol{k};\boldsymbol{\beta}\rangle$.  The dual states of the inertial 
states are called the spacetime states and they are represented by the vector $\vert\xi\rangle$, 
or $\vert\tau,\boldsymbol{\xi}\rangle$, in a rest-frame of an observer. 

The states of all $m$ and $\boldsymbol{k}$ form a complete basis of the relativistic quantum 
Hilbert space of a particle at rest, and the vector $\vert\xi\rangle$, or $\vert\tau,\boldsymbol{\xi}
\rangle$, form also a complete basis for the spacetime location of the particle (see Section 5).  
The fact that neither $m$ nor $\boldsymbol{k}$ are quantities with set up values (see Section 
6) makes the presented RQM to be a significantly different theory than standard RQM, and 
its physical implications are far reaching (see Section 7). 
\item The inertial state $\vert m,\boldsymbol{k}\rangle$ is an eigenvector of the operators 
$\mathcal{K}_a$ such that
	\begin{equation}
        \mathcal{K}_a\vert m,\boldsymbol{k}\rangle = k_a\vert m,\boldsymbol{k}\rangle\ ,
        \label{S3eq5}
	\end{equation}
with the eigenvalues $k=\{m,\boldsymbol{k}\}$. It follows that $k_a k^a=m^2-\vert\boldsymbol{k}
\vert^2 = m^2-\tilde{m}^2 =M^2$. It must be also noted that the vector-mass is completely 
independent from the scalar-mass.
\item The wavefunction of a quantum particle at rest, with the scalar-mass $m$ and vector-mass 
$\boldsymbol{k}$, expressed in the spacetime basis is given by 
	\begin{equation}
        \langle\tau,\boldsymbol{\xi}\vert m,\boldsymbol{k}\rangle=\frac{e^{-i\left(m\tau-
        \boldsymbol{k}\cdot\boldsymbol{\xi}\right)}}{(2\pi)^2}\ ,
        \label{S3eq6}
	\end{equation}
where $\tau=\xi^0$, and the Lorentz-transformed wavefunction of a particle moving with a 
well-defined velocity $\boldsymbol{\beta}$ is given by
    \begin{equation}
        \langle\tau,\boldsymbol{\xi}\vert m,\boldsymbol{k};\boldsymbol{\beta}\rangle=
        \frac{e^{-i\tau(E+\boldsymbol{\beta}\cdot\boldsymbol{V})}e^{i\boldsymbol{\xi}
        \cdot(\boldsymbol{V}+\boldsymbol{\beta}E)}}{(2\pi)^2}\ .
        \label{S3eq7}
    \end{equation}
The state in Eq. (\ref{S3eq6}) describes a particle at rest, nevertheless, the wavefunction seems 
to be propagating in the direction of $\boldsymbol{k}$.  However, a more careful look shows that 
it is not propagating, but instead it is the winding of the complex wavefunction.   Thus, there is 
nothing actually moving though space and the only way to assign motion to the particle is to change 
the motion of the inertial frame of the observer relative to the frame of the particle through a Lorentz 
boost, as given in Eq. (\ref{S3eq7}).
\item The timelike differential operator $Q_t=t^a \mathcal{K}_a$, where $t^a$ is a timelike 
vector, is the generator of translation in time of the relativistic wavefunction.  However, the 
spacelike operator $Q_s=s^a\mathcal{K}_a$, where $s^a$ is spacelike and is orthogonal
to $t^a$, is the translation generator in space.  The symmetries generated by these generators 
correspond to conservation laws, in which the energy and momentum conservation laws are 
unified into one, which is the conservation of scalar-mass energy generated by the timelike 
operator.   For the spacelike operator, the theory suggests that there is also another conservation 
law that pertains to the vector-mass energy, and its physical implications are discussed in Section 7.
\end{enumerate}
\noindent
The above postulates form the basis of our new RQM, which is developed here for a massive 
and spin-zero elementary particle uniquely described by the scalar wavefunction $\phi$ for 
which we now obtain its wave equation.

\section{Generalized Klein-Gordon equation and its solutions}

The eigenvalue equations (see Eq. \ref{S3eq5}), in general, satisfy the following 
second-order Klein-Gordon (KG) eigenvalue equation
    \begin{equation}
        \mathcal{K}_a\mathcal{K}^a\phi\left(\tau,\boldsymbol{\xi}\right)=
        \left(m^2-\vert\boldsymbol{k}\vert^2\right)\phi\left(\tau,\boldsymbol{\xi}\right)=
        M^2\phi\left(\tau,\boldsymbol{\xi}\right)\ ,
        \label{S4eq1}
    \end{equation}
where $\phi=\langle\tau,\boldsymbol{\xi}\vert\phi\rangle$ and $M^2$, whose value ranges 
from $-\infty$ to $\infty$, is the eigenvalue of the operator $\mathcal{K}_a\mathcal{K}^a$.
The general solutions to Eq. {\ref{S4eq1}} for a particular $M^2$ is given as
    \begin{eqnarray}
        \phi \left(\tau,\boldsymbol{\xi}\right) &=& A e^{-i K_a\xi^a}+Be^{-i\bar{K}_a\xi^a} \nonumber\\
        &=& \left[Ae^{-iE(\tau-\boldsymbol{\beta}\cdot\boldsymbol{\xi})}+
        Be^{iE(\tau-\boldsymbol{\beta}\cdot\boldsymbol{\xi})}\right]
        e^{i\boldsymbol{V}\cdot(\boldsymbol{\xi}-\boldsymbol{\beta}\tau)}\ ,
        \label{S4eq3}
    \end{eqnarray}
where $K=P+V$ for the positive energy solution, and $\bar{K}=-P+V$ for the negative 
energy solution, which corresponds to $m\to -m$.  The coefficients $A$ and $B$ are 
constants of integration.  Note that $M^2$ is not particle's mass despite its place in the 
KG equation.  The main reason is that the particle's mass must be in a superposition of 
different vector-mass $\boldsymbol{k}$, which does not play any role in the motion of 
the particle because of its spacelike origin.  

It must be noted that Eq. {\ref{S4eq1}} is not a dynamical equation like the other wave 
equations; in fact, it is the extended form of the Helmholtz equation with the Minkowski 
metric, where the eigenvalue $M^2$ cannot be replaced by an operator without 
introducing new structures to the particle or spacetime, such as internal symmetries or 
extra dimensions, respectively.  Therefore, accounting for spins and different generations
of particles is still required, which is beyond the scope of this paper, as the main focus 
herein is to develop an alternative theory of RQM and present its physical implications.  
In the following, we discuss the relation between the completeness of Hilbert space and 
the causal structure of this new RQM in detail.

\section{Completeness of Hilbert space and causality}

After defining the basic postulates of the new RQM and presenting the resulting generalized 
Klein-Gordon eigenvalue equation, the completness of Hilbert space and causality must be 
considered.  This requires the generalization of the concept of the inner product of the new 
state vectors in Hilbert space, which is defined as
    \begin{equation}
        \langle\tau',\boldsymbol{\xi}'\vert\tau,\boldsymbol{\xi}
        \rangle = \delta(\tau'-\tau)\delta^3\left(\boldsymbol{\xi}'-
        \boldsymbol{\xi}\right)\ ,
        \label{S5eq1}
    \end{equation}
and
    \begin{equation}
        \langle m',\boldsymbol{k}'\vert m,\boldsymbol{k}
        \rangle = \delta(m'-m)\delta^3\left(\boldsymbol{k}'
        -\boldsymbol{k}\right)\ .
        \label{S5eq2}
    \end{equation}
It must be noted again that a spacetime state $\vert\tau,\boldsymbol{\xi}\rangle$ 
is completely defined throughout spacetime, which describes a particle that exists 
exactly at the point $\boldsymbol{\xi}$ in space and $\tau$ in time of a local inertial 
observer.  For such state, an observer would see the particle suddenly come in and 
out of existence, with undefined mass, energy, and momentum.   Conversely, an 
inertial quantum state $\vert m,\boldsymbol{k};\boldsymbol{\beta}\rangle$ 
represents a particle of well-defined scalar-mass, vector-mass, and velocity, but it 
does not have a well-defined spacetime position.

Using Eq. (\ref{S3eq7}), we write the projection of a inertial state onto a spacetime 
state as 
    \begin{equation}
        \langle\tau,\boldsymbol{\xi}\vert m,\boldsymbol{k};\boldsymbol{\beta}\rangle 
        = \frac{e^{-i E\left(\tau-\boldsymbol{\beta}\cdot\boldsymbol{\xi}\right)}
        e^{i\boldsymbol{V}\cdot\left(\boldsymbol{\xi}-\boldsymbol{\beta}\tau\right)}}
        {(2\pi)^2}=\frac{e^{-i\tau(E+\boldsymbol{\beta}\cdot\boldsymbol{V})}
        e^{i\boldsymbol{\xi}\cdot(\boldsymbol{V}+\boldsymbol{\beta}E)}}{(2\pi)^2}\ ,
        \label{S5eq3}
    \end{equation}
which is a state of definite $K$ in flat spacetime, with $\boldsymbol{\beta}=
\boldsymbol{P}/E$ and $\tau=\xi^0$. Since the identity operator $\mathcal{I}$ in the 
spacetime basis is given by
    \begin{equation}
        \mathcal{I}=\int\mathrm{d}^4\xi\ \vert\tau,\boldsymbol{\xi}\rangle\langle\tau,
        \boldsymbol{\xi}\vert\ ,
        \label{S5eq4}
    \end{equation}
we use Eq. (\ref{S5eq3}) to find 
    \begin{eqnarray}
        \langle m',\boldsymbol{k}';\boldsymbol{\beta}\vert\mathcal{I}
        \vert m,\boldsymbol{k};\boldsymbol{\beta}\rangle&=&\int\frac{\mathrm{d}^4\xi}
        {(2\pi)^4}\ e^{i\tau(\Delta E+\boldsymbol{\beta}\cdot\Delta\boldsymbol{V})}
        e^{-i\boldsymbol{\xi}\cdot(\Delta\boldsymbol{V}+\boldsymbol{\beta}\Delta E)}\nonumber\\
        &=&\delta(\Delta E+\boldsymbol{\beta}\cdot\Delta\boldsymbol{V})\delta^3
        (\Delta\boldsymbol{V}+\boldsymbol{\beta}\Delta E)\ ,
        \label{S5eq5}
    \end{eqnarray}
where $\Delta K=K'-K$.  Note that for all allowed values of $\vert\boldsymbol{\beta}\vert<1$, 
Eq. (\ref{S5eq5}) is equivalent to Eq. (\ref{S5eq2}), and that the identity operator $\mathcal{I}$ 
is complete and the boost operator $\mathcal{U}_{\boldsymbol{\beta}}$ (see Eq. \ref{S3eq4}) 
is unitary. Equivalently, the identity operator can be expressed in the inertial basis as
    \begin{equation}
        \mathcal{I}=\int\mathrm{d}m\ \mathrm{d}^3\boldsymbol{k}
        \vert m,\boldsymbol{k}\rangle\langle m,\boldsymbol{k}\vert\ ,
        \label{S5eq6}
    \end{equation}
which is to show that the inertial states are also complete.

Next, we introduce two new operators, $\mathcal{T}_{\boldsymbol{\beta}}$ and 
$\mathcal{S}$, whose definitions are motivated by their application in the 
next section for the construction of the path integral formulation of the new RQM. 
The operators are defined as
    \begin{equation}
        \mathcal{T}_{\boldsymbol{\beta}}(m)=\int\mathrm{d}^3\boldsymbol{k}\ 
        \vert m,\boldsymbol{k};\boldsymbol{\beta}\rangle\langle m,\boldsymbol{k};
        \boldsymbol{\beta}\vert =\int\frac{\mathrm{d}^3\boldsymbol{V}}{\gamma}\ 
        \vert m,\boldsymbol{k};\boldsymbol{\beta}\rangle\langle m,\boldsymbol{k};
        \boldsymbol{\beta}\vert\ ,
        \label{S5eq7}
    \end{equation}
and
    \begin{equation}
        \mathcal{S}(\tau)=\int\mathrm{d}^3\boldsymbol{\xi}\ 
        \vert\tau,\boldsymbol{\xi}
        \rangle\langle\tau,\boldsymbol{\xi}\vert\ .
        \label{S5eq8}
    \end{equation}
To understand the properties of these operators, we investigate the matrix elements in their 
respective dual basis.  First, we study the properties of $\mathcal{T}_{\boldsymbol{\beta}}$ 
with its matrix element given by
    \begin{eqnarray}
        \langle\tau',\boldsymbol{\xi}'\vert\mathcal{T}_{\boldsymbol{\beta}}
        \vert\tau,\boldsymbol{\xi}\rangle&=&\frac{e^{-i E\left(\Delta\tau-\boldsymbol{\beta}
        \cdot\Delta\boldsymbol{\xi}\right)}}{2\pi\gamma}\int\frac{\mathrm{d}^3
        \boldsymbol{V}}{(2\pi)^3}\ e^{i\boldsymbol{V}\cdot\left(\Delta\boldsymbol{\xi}-
        \boldsymbol{\beta}\Delta\tau\right)}\nonumber\\&=&\frac{e^{-i E\left(\Delta\tau-\boldsymbol{\beta}
        \cdot\Delta\boldsymbol{\xi}\right)}}{2\pi\gamma}\ \delta^3\left(\Delta\boldsymbol{\xi}
        -\boldsymbol{\beta}\Delta\tau\right)\ .
        \label{S5eq9}
    \end{eqnarray}
Note that we did not integrate over the scalar-mass $m$. This operator is defined for a 
particle of definite scalar-mass. The immediate interpretation of this result is that this is 
exactly the causal path of the particle traveling between the two spacetime states, where 
$\Delta\xi = \xi'-\xi$. For a particle of definite mass $m$, the result shows that the
particle must be stationary in space when its velocity $\boldsymbol{\beta}=0$, 
while the complex phase evolves over proper time of the particle, i.e.
    \begin{equation}
        \langle\tau',\boldsymbol{\xi}'\vert\mathcal{T}_{\boldsymbol{\beta}=\boldsymbol{0}}
        \vert\tau,\boldsymbol{\xi}\rangle=\frac{e^{-i m\Delta\tau}}{2\pi}\ 
        \delta^3\left(\Delta\boldsymbol{\xi}\right)\ .
        \label{S5eq10}
    \end{equation}

Alternatively, when $\boldsymbol{\beta}\neq 0$, the result (\ref{S5eq9}) is nonzero if, 
and only if, the states are causally separated, i.e. the path traced out by the particle must 
stay within the light cone.  The complex phase difference between the final and initial 
states is given by the complex exponential prefactor. Remarkably, this is simply the 
quantum path of a free point particle and the matrix element is just the phase evolution 
along the path of propagation. This operator is the propagation operator for a particle 
of mass $m$ moving from $\vert\tau,\boldsymbol{\xi}\rangle$ to $\vert\tau',
\boldsymbol{\xi}'\rangle$; however, it is different from time evolution operator 
in QM because the time dependence is already contained in the spacetime states 
$\vert\tau,\boldsymbol{\xi}\rangle$, while the operator $\mathcal{T}_{\boldsymbol
{\beta}}$ simply connects the two spacetime states by imparting the particle with 
a particular mass $m$ moving at the velocity $\boldsymbol{\beta}$.  

In fact, the particle itself can be associated with the operator $\mathcal{T}_{\boldsymbol
{\beta}}$. However, as we recall the general solution to the KG eigenvalue equation in Eq. 
(\ref{S4eq3}), we see that it is a linear combination of both positive and negative energy terms; 
therefore, we must also form a linear combination of the operator $\mathcal{T}_{\boldsymbol
{\beta}}$ with its negative energy counter part, namely $m\to-m$, to describe a particle of 
mass $m>0$. 

Let us refer to the linear combination 
    \begin{equation}
        \mathcal{P}_{\boldsymbol{\beta}}(m)=
        a^{+}_{\boldsymbol{\beta}}\mathcal{T}_{\boldsymbol{\beta}}(m)+
        b^{-}_{\boldsymbol{\beta}}\mathcal{T}_{\boldsymbol{\beta}}(-m)\ ,
        \label{S5eq11}
    \end{equation}
as the 'particle operator' that creates a particle or destroys an antiparticle of mass $m$, 
and its corresponding mass conjugation 
    \begin{equation}
        \mathcal{P}^{\star}_{\boldsymbol{\beta}}(m)=
        a^{-}_{\boldsymbol{\beta}}\mathcal{T}_{\boldsymbol{\beta}}(-m)+
        b^{+}_{\boldsymbol{\beta}}\mathcal{T}_{\boldsymbol{\beta}}(m)\ ,
        \label{S5eq12}
    \end{equation}
as the 'antiparticle operator' that creates an antiparticle or destroys a particle of mass $m$. 

Here, we have promoted the numerical coefficients to creation and annihilation operators 
that act on Fock space of occupancy number basis, $\vert n\rangle$, and defined the mass 
conjugation of the operator $\mathcal{T}_{\boldsymbol{\beta}}$ as
    \begin{equation}
        \mathcal{T}^{\star}_{\boldsymbol{\beta}}(m)=\mathcal{T}_{\boldsymbol{\beta}}(-m)\ ,
        \label{S5eq13}
    \end{equation}
and
    \begin{equation}
        (a^{+}_{\boldsymbol{\beta}})^{\star}=a^{-}_{\boldsymbol{\beta}}\ .
    \end{equation}

By promoting the coefficients to operators, the spacetime states must also be promoted to the 
spacetime Fock's states $\vert\tau,\boldsymbol{\xi};n\rangle=\vert\tau,\boldsymbol{\xi}\rangle
\otimes\vert n\rangle$, such that $\vert\tau,\boldsymbol{\xi};0\rangle$ represents the vacuum 
state and $\vert\tau,\boldsymbol{\xi}\rangle$ represents single particle state.  Thus, we have
    \begin{equation}
        \langle\tau',\boldsymbol{\xi}'\vert\mathcal{P}_{\boldsymbol{\beta}}\vert\tau,\boldsymbol{\xi};0\rangle
        =\langle\tau',\boldsymbol{\xi}'\vert\mathcal{T}_{\boldsymbol{\beta}}\vert\tau,\boldsymbol{\xi}\rangle\ .
        \label{S5eq14}
    \end{equation}

Finally, we have the matrix element of the operator $\mathcal{S}$ expressed as follows
    \begin{equation}
        \langle m',\boldsymbol{k}';\boldsymbol{\beta}\vert
        \mathcal{S}\vert m,\boldsymbol{k};\boldsymbol{\beta}\rangle=
        \frac{e^{-i\tau(\Delta E+\boldsymbol{\beta}\cdot\Delta\boldsymbol{V})}}
        {2\pi}\delta^3\left(\Delta\boldsymbol{V}+\boldsymbol{\beta}\Delta E\right)\ ,
        \label{S5eq15}
    \end{equation}
which is the spatial inner product of two inertial states at time $\tau$.  The $\mathcal{S}$ operator 
specifies a local inertial frame, in which the inner product can be calculated, or the spatial projection 
of one particle state onto another.  As shown by Eq. (\ref{S5eq15}), the projection is given for two 
states of different four-mass vectors with the same velocity.  This may be interpreted as the transition 
amplitude of a particle changing its mass as it propagates.  In another case, we could project states 
of the same masses but different velocities. The result would be the transition amplitude of the particle
changing its velocity.  In either case, the particle is undergoing an interaction, where the former could 
describe the particle decay process and the latter for a particle experiencing a force.  However, in this 
paper we consider only free particles.

In the next section, the pair $\mathcal{S}$ and $\mathcal{T}_{\boldsymbol{\beta}}$ will be 
used to recover the relativistic classical Lagrangian of a free particle by following the Feynman 
path integral formulation of QM.   Our goal is to obtain a consistent relativistic path integral 
and, ultimately, confirm the validity of our reformulated theory of RQM.

\section{Path integral formulation}

We now present the path integral formulation of the developed RQM by deriving the transition 
amplitude for a particle with a definite scalar-mass $m$. The integration is performed over all 
possible $\vert\boldsymbol{\beta}\vert<1$, or equivalently,  $\vert\gamma\boldsymbol{\beta}
\vert<\infty$, since a quantum particle can take any path and the velocity integration is bounded 
because the particle is massive.  Then, using the operator $\mathcal{T}_{\boldsymbol{\beta}}$
from previous section (see Eq. \ref{S5eq7}), the general transition operator $\mathcal{T}$ is 
given by
    \begin{equation}
        \mathcal{T}(m)=\int_{\vert\gamma\boldsymbol{\beta}\vert<\infty}\frac{\mathrm{d}^3
        (\gamma\boldsymbol{\beta})}{\gamma}\ \mathcal{T}_{\boldsymbol{\beta}}(m)=
        \int_{\vert\boldsymbol{\beta}\vert<1}\mathrm{d}^3\boldsymbol{\beta}\ \gamma^2
        \mathcal{T}_{\boldsymbol{\beta}}(m)\ .
        \label{S6eq1}
    \end{equation}
The presence of the factor $\gamma$ is to keep the integration Lorentz-invariant.  Using 
the particle operator to create a particle of mass $m$ out of the vacuum as shown in Eq. 
(\ref{S5eq14}), we integrate Eq. (\ref{S5eq9}) to obtain the relativistic quantum transition 
amplitude 
    \begin{eqnarray}
        \langle \tau',\boldsymbol{\xi}'\vert\mathcal{T}\vert\tau,
        \boldsymbol{\xi}\rangle&=&\int_{\vert\boldsymbol{\beta}\vert<1}\mathrm
        {d}^3\boldsymbol{\beta}\ \frac{e^{-iE\left(\Delta\tau-\boldsymbol{\beta}
        \cdot\Delta\boldsymbol{\xi}\right)}}{2\pi\gamma^{-1}}\ \delta^3
        \left(\Delta\boldsymbol{\xi}-\boldsymbol{\beta}\Delta\tau\right)\nonumber\\&=&
        \frac{e^{-im\Delta\bar{\tau}}}{2\pi\left\vert\Delta\bar{\tau}\right\vert}\ 
        \Theta\left(1-\left\vert\frac{\Delta\boldsymbol{\xi}}{\Delta\tau}\right\vert^2\right)\ ,
        \label{S6eq2}
    \end{eqnarray}
where
    \begin{equation}
        \Delta\bar{\tau}=\Delta\tau\sqrt{1-\left\vert\frac{\Delta\boldsymbol{\xi}}
        {\Delta\tau}\right\vert^2}\ ,
        \label{S6eq3}
    \end{equation}
and the function $\Theta(x)$ is the Heaviside step function, whose presence 
is necessary to prevent the particle from propagating faster than the speed of light. 

The Feynman path integral formalism describes the procedure of calculating the propagator 
matrix element $G(\xi_F;\xi_I)$, a.k.a Green's function, of a free particle moving from $\xi$ 
to $\xi'$. This is done by integrating the Feynman propagation phase factor over all possible 
timelike paths that the particle could take.  To obtain this operator, firstly we subdivide 
spacetime phase space into $N$ slices of spacelike hypersurfaces $\Sigma_n$ and the 
corresponding dual surfaces $\Omega_n$, where $\boldsymbol{\xi}_n\in\Sigma_n$ 
and $\boldsymbol{k}_n\in\Omega_n$.  Each surface $\Sigma_n$ is spatial surface that 
represents a time slice in spacetime where the particle could be found at time $\tau_n$. 
The integration over this surface is represented by the operator $\mathcal{S}_{\tau_n}$. 

Then, we connect the different hypersurfaces using the summation over all world lines 
that the particle could take. This is done by including the transition operators 
$\mathcal{T}_{n}$ of a fixed scalar-mass in between $\mathcal{S}_{\tau_n}$. 
Lastly, we take the continuum limit as $N \rightarrow \infty$ and $\Delta\tau_n 
\rightarrow 0$. This construction gives us the following the time-ordered products
    \begin{eqnarray}
        G_N(\xi_F,\xi_I) &=& \langle\xi_F\vert\mathcal{U}_N\vert\xi_I\rangle \nonumber\\
        &=& \langle\xi_N
        \vert \mathcal{T}_{N}C_{N-1}\mathcal{S}_{\tau_{N-1}}\mathcal{T}_{{N-1}}
        \cdots C_{n}\mathcal{S}_{\tau_n}\mathcal{T}_{n}\cdots C_1\mathcal{S}_{\tau_1}
        \mathcal{T}_{1}\vert\xi_0\rangle \nonumber\\
        &=& \left[\prod_{n=1}^{N-1}\int_{\Sigma_{n}}\mathrm{d}^3
        \boldsymbol{\xi}_{n}C_n\right]\left[\prod_{j=1}^N\langle\xi_j
        \vert\mathcal{T}_{j}\vert\xi_{j-1}\rangle\right]\ ,
        \label{S6eq4}
    \end{eqnarray}
where $\tau_n>\tau_{n-1}$ and $C_n$ is some normalization constant to be determined. The 
matrix elements of $\mathcal{T}_{j}$ are given by Eq. (\ref{S6eq2}). Thus, we have
    \begin{equation}
        \langle\xi_j\vert\mathcal{T}_{j}\vert\xi_{j-1}\rangle=\frac{e^{-im\Delta\bar{\tau}_j}}
        {2\pi\Delta\bar{\tau}_j}\Theta\left(1-\left\vert\frac{\Delta\boldsymbol{\xi}_j}{\Delta\tau_j}
        \right\vert^2\right)\ ,
        \label{S6eq5}
    \end{equation}
which leads to
    \begin{equation}
        G_N(\xi_F,\xi_I)=\left[\prod_{n=1}^{N-1}\int\mathrm{d}^3\boldsymbol{\xi}_{n}C_n\right]
        \prod_{j=1}^N\left(\frac{e^{-im\Delta\bar{\tau}_j}}{2\pi\Delta\bar{\tau}_j}\right)\ ,
        \label{S6eq6}
    \end{equation}
where $\Delta\xi_j=\xi_j-\xi_{j-1}$. Also, we have omitted the Heaviside functions for legibility and
    \begin{equation}
        \Delta\bar{\tau}_j=\Delta\tau_j\sqrt{1-\left\vert\frac{\Delta\boldsymbol{\xi}_j}{\Delta\tau_j}
        \right\vert^2}\ .
        \label{S6eq7}
    \end{equation}

With the endpoints fixed and the velocity bounded by the speed of light, we take the limit 
$N\to\infty$ and obtain
    \begin{equation}
        G(\xi_F,\xi_I)=\int\mathrm{D}\boldsymbol{\xi}(\bar{\tau})\ Ce^{i S}\ ,
        \label{S6eq8}
    \end{equation}
where the integral is appropriately normalized by $C$.  Then, the path-integral measure 
in the continuum limit is
    \begin{equation}
        \mathrm{D}\boldsymbol{\xi}(\bar{\tau})=\lim_{N\to\infty}\prod_{n=1}^{N-1}
        \mathrm{d}^3\boldsymbol{\xi}_{n}\ .
        \label{S6eq9}
    \end{equation}

The above results demonstrate that the classical action of a relativistic free particle is correctly 
recovered from the path-integral formulation, with the action given by 
    \begin{equation}
        S=-m\int_{\bar{\tau}_I}^{\bar{\tau}_F}\mathrm{d}\bar{\tau}\ ,
        \label{S6eq10}
    \end{equation}
where $\bar{\tau}$ is the proper time of a relativistic particle.  We may finally define the 
evolution operator $\mathcal{U}$ that acts on states in spacetime basis as
    \begin{equation}
        \mathcal{U}(\tau)=\lim_{N\to\infty}\mathcal{T}_{N}\prod_{n=1}^{N-1}C_n\mathcal{S}_{\tau_n}
        \mathcal{T}_{n}\ .
        \label{S6eq11}
    \end{equation}

Now, developing an analytical method to evaluate the relativistic path integral given in Eq. 
(\ref{S6eq8}) is out the scope of this paper.  Nonetheless, for a free particle this problem 
is equivalent to showing that it reduces to the result (see Eq. \ref{S6eq2}) for the operator 
$\mathcal{T}$.
\begin{table}
\caption{Differences between the standard and new RQM}\label{tab1}
\centering
\begin{tabular}{| l | c | c | c }
\hline
\textbf{Concepts} & \textbf{Standard RQM}  & \textbf{New RQM}\\
\hline
Four momentum & Timelike & Timelike and spacelike\\
Invariant mass & Scalar & Scalar and vector\\
Particle's velocity& Implicit & Explicit\\
Conjugate variables & Energy - time & Scalar-mass - proper time\\
Conjugate variables & Momentum - position & Vector-mass - proper distance\\
Conserved quantities & Timelike components & Timelike and spacelike components\\
Path integral & All trajectories & Trajectories inside the light cone\\ 
\hline
\end{tabular}
\end{table}

\section{Physical implications}

The main differences between the standard RQM and the reformulated RQM presented 
in this paper are summarized in Table 1.  Let us now briefly comment on the concepts 
given in Table 1.  As a consequence of symmetry between the timelike and spacelike 
intervals, the four momentum is no longer confined to the light cone, and that the 
theory requires both the scalar-mass and vector-mass; although only scalar-mass is 
observable, the four-mass vector labels the mass state of the particle.  The (linear) 
velocity is a parameter of the Lorentz boost.  The reformulation makes explicit 
distinction between the vector-mass and the velocity vector.  The quantum nature 
of the particle is encoded in the superposition of different vector-masses and velocities. 
The proper time and distance are defined relative to an inertial observer and the 
reformulation redefines the Heisenberg uncertainty principles with the new conjugate 
variable pairs.  Furthermore, the generalized KG equation obtained in this paper is not 
a wave equation but an eigenvalue equation. Unlike the square of the scalar-mass in 
the conventional theory being a constant, $M^2$ is an eigenvalue.  On the one hand, 
the conservation of timelike component incorporates both the conservation of momentum 
and energy; on the other hand, the conservation of spacelike component could be related 
to the conservation of quantum information and responsible for the nonlocal quantum 
effects. Most importantly, the reformulated theory reproduces the well-known relativistic 
Lagrangian of a classical particle for the path integral formulation.

The reformulated RQM presented in this paper takes into account the spacelike eigenvalues 
of the linear momentum, in addition to its timelike eigenvalues.  The timelike and spacelike 
eigenvalues are associated with the scalar-mass $m$ and vector-mass $\boldsymbol{k}$, 
respectively, with the latter being a new physical concept.  The theory demands that $m$ 
and vector-mass $\boldsymbol{k}$ are independent observable quantities.  However, in 
practice only $m$ can be observed since all observers and their experiments are confined 
to their local light cone.  On the other hand, $\boldsymbol{k}$ does not describe the 
particle's causal motion to an inertial observer, therefore, it cannot be measured 
experimentally because of its spacelike nature outside of the light cone.  Despite this 
spacelike nature of $\boldsymbol{k}$, the developed theory is not a hidden-variable
theory and it preserves causality.  

Moreover, the presence of vector-mass does not affect particle's probability amplitude
because $\boldsymbol{k}$ is integrated out of the complex phase when the operator 
$\mathcal{T}_{\boldsymbol{\beta}}$ is projected onto the spacetime states.  The 
result is that the complex phase that involves $\boldsymbol{k}$ becomes a delta 
function representing the localization of the particle in spacetime.  The remaining 
complex phase contains the scalar-mass and the proper time, which becomes the 
probability amplitude of the particle for a particular path that it takes as it is shown 
by the path integral formulation (see Section 6).   Other important physical implications 
of the presence of the vector-mass in the theory include nonlocality, a novel picture 
of vacuum, and some modifications of the uncertainty principle, and they are now 
discussed.

There is a large body of literature on the subject of nonlocality and incompleteness in
QM and the subsequent developments triggered by the original Einstein, Podolsky and 
Rosen paper [36] and Bell's work [37-39].  Bell showed that hidden-variables theories 
must be nonlocal, in order to be in agreement with the empirically verified predictions 
of QM, and that {\it nonlocality} of QM cannot be attributed to incompletness [40].
The work on nonlocality in QM initiated formulations of different nonlocal theories 
of QFT.  In the earlier work summarized in [41], the nonlocal theories were constructed 
by either introducing a new universal length that allows for a slight violation of locality 
on small distances, or defining a constant that characterizes the domain of nonlocal 
interactions.  In more recent work, nonlocal QFT for scalar fields were constructed 
[42,43].  

In the RQM theory presented in this paper, the scalar-mass is responsible for the 
time evolution of the wavefunction, similar to standard RQM; however, the 
vector-mass affects only the spatial distribution of the wavefunction, which 
makes our theory a fully symmetric theory and, thus, different than those
local and nonlocal theories previously constructed [41-43].  We identify the 
spacelike nature of the wavefunction spatial distribution with {\it nonlocality}
in our theory and suggest that due to the presence of vector-mass, the particle 
may potentially interact nonlocally with another particle by overlapping the
spatial distributions of their wavefunction.  Since our theory considers only one 
free, massive and spin zero elementary particle, such nonlocally interacting 
particles are out of the scope of this paper.

In this reformulated RQM, the scalar-mass is responsible for the the particle's
world-line being within the light cone.  Moreover, a particle of definite mass and 
position in space must be in a superposition of different vector-masses, which 
means that $\boldsymbol{k}$ does not play any role in the motion of the particle.
As a result, $M^2$ given by Eq. (\ref{S2eq4}) is not the mass of the particle and
the theory allows only for particles with positive $m^2$ term, which prevents 
tachyons and tachyonic fields [31,32] from being defined in this theory.

The vector-mass, as the spacelike component responsible for the nonlocal behavior 
of quantum particles, is the true conjugate variable of position in $\mathbb{R}^3$ 
space, thus, the canonical commutation relations given by Eq. (\ref{S3eq1}) can be
used to write the uncertainty principle in the following form
    \begin{equation}
	\prod_{i=1}^3 \Delta \xi^{i}\Delta k^{i} > 1\ .
       \label{S7eq1}
	\end{equation}
There is also uncertainty of the scalar-mass due to the presence of $\boldsymbol{k}$
and this uncertainty is inversely proportional to particle's lifetime, then the resulting 
uncertainty relation becomes 
	\begin{equation}
	\Delta\tau\Delta m > 1\ ,
       \label{S7eq2}
	\end{equation}
and it describes virtual particles that are present in this reformulated RQM; note that 
according to our theory only particles with infinite lifetime are stable and observable.
The presence of virtual particles in the theory leads to a novel picture of vacuum 
that is now briefly described.

From a point of view of RQM and QFT, vacuum is defined as the state of lowest energy
filled with virtual particles spontaneously coming out of the void and disappearing in it
(e.g., [44,45]), with the notion that known elementary particles may form pairs with 
their antiparticles and become virtual on the time and energy scales corresponding 
to these particles (e.g., [46-48]).  The theory presented in this paper and its uncertainty
relation given by Eq. (\ref{S7eq2}) shows masses of virtual particles do not have to be 
limited by experimentally established masses of known elementary particles, as virtual
particles may have any mass, including masses of all known particles, which are 
required by this theory to be stable and detectable if, and only if, their lifetimes are 
infinititely long. 

\section{Conclusions}

Standard RQM that is based on the timelike intervals is reformulated by taking 
into account both the timelike and spacelike intervals.  The presence of the 
spacelike intervals in the reformulated RQM modifies its linear momentum
by including its spacelike eigenvalues, in addition to timelike eigenvalues. 
The theory is developed for a free, massive, and spin zero elementary 
particle in the Minkowski spacetime, where the particle is called elementary
if its wavefunction transforms as one of the irreps of the Poincar\'e group. 
The STR energy-momentum relationship of standard RQM with its scalar-mass 
is modified by an additional term with its vector-mass, which only affects the 
spatial distribution of particle's wavefunction.  The physical consequences of 
the presence of the vector-mass in the energy-momentum relationship are far 
reaching, starting with the modification fo the mass term in the Klein-Gordon 
equation, and then leading to nonlocality, some modifications of the uncertainty 
principle, and a novel picture of vacuum.  Moreover, the developed RQM theory 
preserves causality, is self-consistent, and is formulated using the path integral 
formalism, which correctly reproduces the classical action of a relativistic particle; 
however, the theory does not allow for tachyons and tachyonic fields.  The theory 
is not a hidden-variable theory; its physical interpretation is consistent with the 
Copenhagen interpretation of QM, and preserves causality.

\bigskip\noindent
{\bf Data Availability}
The data that supports the findings of this study is available within the article.

\bigskip\noindent
{\bf Acknowledgments}
We are grateful to two anomynous referees for their comments and suggestions
that allowed us to improved the original version of this paper.  We also thank 
Lesley Vestal for reading our manuscript and commenting on it.
%

\end{document}